\documentclass[12pt]{article}
\usepackage{amsmath,amsfonts, amssymb, braket, bbold}
\usepackage{tikz}
\usetikzlibrary{trees,er,snakes,shapes,mindmap}
\usepackage{titlesec}
\titleformat{\section}
 {\normalfont\fontfamily{put}\fontsize{12pt}{16pt}\bfseries\color{black}}
{\thesection}{1em}{}
\titleformat{\subsection}
 {\normalfont\fontfamily{put}\fontsize{12pt}{16pt}\bfseries\color{black}}
{\thesubsection}{1em}{}
\def \beq  {\begin{equation}}
\def \eeq  {\end{equation}}
\def \beqar {\begin{eqnarray}}
\def \eeqar {\end{eqnarray}}
\allowdisplaybreaks
\def\sqr#1#2{{\vcenter{\vbox{\hrule height.#2pt
\hbox{\vrule width.#2pt height#1pt \kern#1pt
\vrule width.#2pt}\hrule height.#2pt}}}}

\def\la {{\langle}}
\def\ra {{\rangle}}

\def\vf {{\varphi}}

\def\Tr {{\rm Tr}}

\def\del {\partial}
\def\bdel{\bar{\partial}}

\def\e {\epsilon}

\def\bz {{\bar{z}}}

\def\C {{\cal C}}
\def\D {{\cal D}}

\def\F {{\cal F}}

\def\H {{\cal H}}
\def\I {{\cal I}}

\def\M{{\cal M}}
\def\O {{\cal O}}

\def\vf {{\varphi}}

\def\bpsi{{\bar\psi}}

\def\half{\textstyle{1\over 2}}
\def \halfn {\textstyle{n\over 2}}

\mathchardef\mhyphen="2D
\begin{document}
%%%%%%%%%%%%%%%%%%%%%%%%%%%%%%%%%%%%%%%%%%%%%%%
%%%%%%%%%%%%%%%%%%%%%%%%%%%%%%%%%%%%%%%%%%%%%%%
%\fontfamily{bch}\fontsize{12pt}{16pt}\selectfont
\fontfamily{put}\fontsize{12pt}{16pt}\selectfont
%\fontfamily{pnb}\fontsize{12pt}{16pt}\selectfont
%\fontfamily{pzc}\fontsize{14pt}{16pt}\selectfont
%\fontfamily{pbk}\fontsize{12pt}{16pt}\selectfont
%\fontfamily{cmr}\fontsize{11pt}{15pt}\selectfont
%\fontfamily{put}\fontsize{12pt}{17pt}\selectfont
%\fontfamily{lmss}\fontsize{11pt}{16pt}\selectfont
%\fontfamily{phv}\fontshape{ro}\fontsize{11pt}{14pt}\selectfont
%\fontfamily{ptm}\fontseries{m}\fontshape{r}\fontsize{12pt}{16pt}\selectfont
%\fontfamily{pnc}\fontseries{m}\fontshape{r}\fontsize{11pt}{15pt}\selectfont
%\fontfamily{ppl}\fontseries{m}\fontshape{r}\fontsize{11pt}{15pt}\selectfont
%\usefont{T1}{phv}{m}{it}
%%%%%%%%%%%%%%%%%%%%%%%%%%%%%%%%%%%%%%%%%%%%%%%
%%%%%%%%%%%%%%%%%%%%%%%%%%%%%%%%%%%%%%%%%%%%%%%
\def \CMP {{Commun. Math. Phys.}}
\def \PRL {{Phys. Rev. Lett.}}
\def \PL {{Phys. Lett.}}
\def \NPBProc {{Nucl. Phys. B (Proc. Suppl.)}}
\def \NP {{Nucl. Phys.}}
\def \RMP {{Rev. Mod. Phys.}}
\def \JGP {{J. Geom. Phys.}}
\def \CQG {{Class. Quant. Grav.}}
\def \MPL {{Mod. Phys. Lett.}}
\def \IJMP {{ Int. J. Mod. Phys.}}
\def \JHEP {{JHEP}}
\def \PR {{Phys. Rev.}}
\def \JMP {{J. Math. Phys.}}
\def \GRG{{Gen. Rel. Grav.}}
%%%%%%%%%%%%%%%%%%%%%%%%%%%%%%%%%%%%%%%%%%%%%%%
%%%%%%%%%%%%%%%%%%%%%%%%%%%%%%%%%%%%%%%%%%%%%%%
\begin{titlepage}
\null\vspace{-62pt} \pagestyle{empty}
\begin{center}
%\rightline{CCNY-HEP-18/4}
%\rightline{August 2018}
\vspace{1.3truein} {\large\bfseries
Remarks on Entanglement for Fuzzy Geometry and Gravity}
\\
{\Large\bfseries ~}\\
\vskip .5in
{\Large\bfseries ~}\\
%%%%%%%%%%%%%%%%%%%%%%%%%%%%%%%%%%%%%%%%%%%%%%%
%%%%%%%%%%%%%%%%%%%%%%%%%%%%%%%%%%%%%%%%%%%%%%%
{\sc V.P. Nair}\\
\vskip .2in
{\sl Physics Department,
City College of New York, CUNY\\
New York, NY 10031}\\
 \vskip .1in
\begin{tabular}{r l}
{\sl E-mail}:&\!\!\!{\fontfamily{cmtt}\fontsize{11pt}{15pt}\selectfont vpnair@ccny.cuny.edu}\\
\end{tabular}
\vskip .5in

%%%%%%%%%%%%%%%%%%%%%%%%%%%%%%%%%%%%%%%%%%%%%%%
%%%%%%%%%%%%%%%%%%%%%%%%%%%%%%%%%%%%%%%%%%%%%%%
\centerline{\large\bf Abstract}
\end{center}
We consider defining a fuzzy space by 
a specific state in a fermionic field theory in terms of which all the observables for the space can be evaluated.
This allows for a definition of entanglement for a fuzzy space by direct integration of the fields over a certain region.
Even though the resulting formula for the entanglement entropy (EE) is
similar to what has been used in the quantum Hall effect,
our derivation provides a novel perspective.
We also review and strengthen the arguments for the EE to be
described by a generalized Chern-Simons form.

\end{titlepage}
%%%%%%%%%%%%%%%%%%%%%%%%%%%%%%%%%%%%%%%%%%%%%%%
%%%%%%%%%%%%%%%%%%%%%%%%%%%%%%%%%%%%%%%%%%%%%%%
\fontfamily{put}\fontsize{12pt}{16pt}\selectfont
\pagestyle{plain} \setcounter{page}{2}
\section{Introduction}
Noncommutative geometry and fuzzy spaces
have been the subject of extensive investigations for many decades by now, with many interesting and beautiful results, primarily of a mathematical nature. (For a general review of this topic, see \cite{reviews}.) Since the key element for such spaces is a Hilbert space
of states, rather than the points of classical geometry, we might consider
these spaces to be ready and primed for a quantum description of spacetime. Consequently, the most significant physical context for which we can expect
these concepts to be appropriate should be quantum gravity.
On this score however, it is fair to say that a convincing scenario has not yet emerged. 

This is not to say that there has not been progress. A possible model for gravity with the inclusion of the Standard Model has been discussed in
\cite{CC}. Many variants of this have also been proposed and analyzed \cite{Lizzi}.
The development of differential geometry for noncommutative (NC) spaces
has been pursued in a number of papers \cite{Aschieri}.
There is also a large body of literature on quantum field theories on a fixed
background which is a noncommutative space \cite{NCfields}.
The matrix model versions of M-theory bring noncommutative geometry
close to string theory and there has been a number of developments along this direction as well \cite{Steinacker}.
Nevertheless, there are still many questions without clear answers.
For example, should time be considered as a noncommuting coordinate
or should it have a different status compared to the spatial coordinates?
While a NC  version of Poincar\'e symmetry can be mathematically 
realized, keeping time as a continuous parameter with a special status,
for example, as the parameter for the modular automorphisms of the
von Neumann algebra of local observables, is also an appealing and
intriguing possibility \cite{rovelli-connes}. The issue of Lorentzian signature versus
the Euclidean one in NC geometry is another point which 
needs clarification, see for example \cite{Lizzi2}.

Rather than continuing with a status report or enumeration of similar issues, 
in this paper, we focus on a couple of questions and new directions which are relevant to relating fuzzy geometry to gravity. We want to characterize the procedures for the limits to continuous geometry
in terms of the metrical and geometric data on the emergent continuous manifold. In other words, this will be a teleological characterization
of the approach to the continuum. Secondly, we want to formulate
fuzzy geometry
in terms of a state (or density matrix)
in the language of quantum field theory.
A prime motivation for this rephrasing is that it will
facilitate formulating the concept of
entanglement for the states defining the manifold itself.
The entanglement entropy (EE) will be related to a generalized Chern-Simons form, bringing gravity on a fuzzy space close to an entropic interpretation.
(It should be emphasized that here we are discussing entanglement 
entropy  due to the degrees of freedom associated with space itself. 
there is some literature on this for certain special cases, but our approach
is different \cite{karcz}.
Entanglement entropy for fields defined on a fuzzy space has been discussed in the literature, see, for example, \cite{ydri} for 
the case of a scalar field.)

Towards a more specific
statement of the problem, we start by recalling that a fuzzy space is defined by a sequence of a triplet
$({\mathcal H}_N , Mat_N , {\Delta}_N )$,
where ${\mathcal H}_N$ is an $N$-dimensional Hilbert space and
$Mat_N$ is the matrix algebra of $N\times N$-matrices which act  
on ${\mathcal H}_N$.
Further, ${ \Delta}_N$ is a  
matrix analog of the Laplacian.
The elements of the sequence are labeled by $N$.
In the limit $N\rightarrow \infty$,
the (noncommutative) matrix algebra $Mat_N$ is well-approximated
by the (commutative) algebra of functions on
a smooth manifold $\M$. The information about the metrical and other geometrical properties of the manifold $\M$ is carried by the operator
${\Delta}_N$.
The limit $N \rightarrow \infty$ is analogous to the passage from a
quantum description of a physical system (in terms of its Hilbert space of states and operators acting on it) to the classical limit (with commutative functions on the phase space $\M$) with $1/N$ playing the role of 
$\hbar$ or the deformation parameter of quantization.

Let us also recall that the purpose of a theory of gravity is to provide a choice of spacetime metric, for example, as the solution of Einstein's equations in
the classical case. In the fuzzy context, since metrical information is encoded in the Laplace operator $\Delta_N$, a theory of gravity must provide a dynamical way of choosing a particular $\Delta_N$,
presumably as dictated by the choice of the state for the matter sector.

A spectral action (based on a Dirac operator rather than $\Delta_N$)
is the usual approach to this question, but
we may note that, in practice, the large $N$-limit of an operator is constructed by using a set of ``wave functions" and then constructing symbols associated
to the operator.
So the first key idea we would like to pursue in this paper is to consider 
the required data for a fuzzy space as $(\H_N, Mat_N)$ along with
a procedure of constructing symbols, rather than directly dealing
with $\Delta_N$.
The manifold $\M$ which emerges in the large $N$ limit will be equipped with a metric (and possibly other geometric structures as well).
This final data will determine the procedure used for constructing symbols.
In other words, we may view the procedure for constructing symbols and the possible resultant large $N$ limits as being parametrized by
the geometric data on the emergent continuum description.

To clarify this further,
we start with the observation that there are two standard ways to associate a function
to an operator ${\hat F}$ on $\H_N$, which facilitates the
large $N$ approximations.
The contravariant symbol
$F$ is related to the matrix elements of ${\hat F}$
(in an orthonormal basis of $\H_N$) as
\beq
F_{r s} = \int_\M d\mu \, u^*_r \, F \, u_s
\label{1}
\eeq
where $F_{r s} = \bra{r} {\hat F}\ket{s}$ are the matrix elements of ${\hat F}$ in the chosen basis for $\H_N$ and $u_r$ are a set of ``wave functions", which are the zero modes of some Laplace operator on $\M$ and $d\mu$ denotes the volume element 
of $\M$. Usually one considers $\M$ to be a complex K\"ahler manifold,
with
the functions $u_r$ obtained as coherent states via geometric
quantization (in the holomorphic polarization) of a multiple of the K\"ahler two-form.
We can regard (\ref{1}) as the Berezin-Toeplitz quantization of the
classical function $F$ on $\M$.
The second way is given by the covariant symbol, which associates the
function $(F)$ to the operator ${\hat F}$ via
\beq
(F) = \C \sum_{r  s} u_r \, F_{r s} \, u^*_s
\label{2}
\eeq
where $\C$ is a constant which depends on the wave functions.
(We will specify $\C$ in more detail below.)
The product of operators lead to a noncommutative
star product for the symbols, which simplifies to the usual commutative product
as $N \rightarrow \infty$, with $\M$ as the emergent continuum
approximation to the fuzzy space. It is clear that the data required
to define $u_r$ is the geometric data on the emergent manifold $\M$.

This is our first key observation: The geometry of the
continuum limits. i.e., the large $N$ limits,
can be labeled by
the choices of $u_r$ 
on the emergent continuum manifold.
Thus the purpose of a theory of gravity
is to provide a way to pick a particular large $N$ limit, which we may consider as ``optimal", equivalent to the choice of the geometric data on $\M$.
This brings us to the second point we want to consider.
It has been known for a while that the Einstein field equations for gravity
 can be viewed as the
extremization of the Bekenstein-Hawking (BH) entropies for Rindler horizons of all accelerated observers in the given spacetime \cite{jacobson}.
(For other related work with entropic interpretation of gravity, see
\cite{verlinde}.)
it is also known that,
at least in some contexts, the BH entropy can be viewed as an entanglement entropy \cite{RT}. This then raises the following possibility:
Rather than considering a spectral action, can we calculate the
entanglement entropy for a fuzzy space as a function of the geometric data
on the emergent continuum manifold (which defines the large $N$ procedure) and obtain the field equations of gravity as the extremization of this entropy?
This is the subject of the present paper.

The rest of this paper is organized as follows.
In the next section, we discuss modeling fuzzy spaces by a 
quantum field theory for the degrees of freedom describing space itself.
In section 3, we use this idea to define a notion of entanglement
for the space, by directly integrating out the degrees of freedom
for the fuzzy space except for those in some region $\O_1 \subset \M$.
This is done using fermionic coherent states and provides a novel
derivation of the result for the entanglement entropy, even though
a similar expression has been used for the Hall effect in the literature.
In section 4, we give a direct calculation of the dependence of the entanglement entropy on 
the background fields in two dimensions using perturbation theory. 
A generalization to higher dimensions is given in section 5.
We argue that the leading term in the
entanglement entropy is described by a generalized Chern-Simons
form expressed in terms of 
the frame fields and the spin connection in the large $N$ limit.
This suggests that the gravitational theory on such spaces in
 the continuum description as $N \rightarrow \infty$
is of the Chern-Simons type.
The paper concludes with a short summary.

\section{A field theoretic state for fuzzy geometry}

We will now specify the fuzzy geometry in terms of a state in a quantum field theory (for the degrees of freedom for space itself).
To provide a clear and concrete framework,
we start with the example of the fuzzy two-sphere; thus
$\M= S^2= {\mathbb{CP}^1} $.
Since this space is also the group coset $SU(2)/U(1)$, we can coordinatize $S^2$ by elements of $SU(2)$, say $g$, with the identification of different $g$'s related by $U(1)$ translations.
Specifically, in a $2\times 2$ matrix representation of $g$, we have the
identification $g \sim ge^{i t_3 \vf}$, where $t_a = \half \sigma_a$, $a =1,2,3$,
and $\sigma_a$ are the Pauli matrices. We define right translation operators
on $g$ by $R_a \, g = g\, t_a$. Spatial derivatives on $S^2$ correspond to
the right translation operators $R_\pm = R_1 \pm i R_2$.
Local complex coordinates $z, \bz$ on $S^2$ are introduced by the parametrization
\beq
g = {1\over \sqrt{1+ \bz z}} \left( \begin{matrix} 
1&z\\
-\bz& 1\\ \end{matrix} \right)
\label{3}
\eeq
with the K\"ahler one-form given by
\beq
a = -{i\over 2} {\bz dz - z d \bz \over (1+ \bz z)}
\label{4}
\eeq
Covariant derivatives on $S^2$ can be identified as
$D_\pm = i R_\pm$. For the wave functions, which we will need for the symbol and the
large $N$ limit, we will use the eigenfunctions of $R_+ R_-$, which is closely related to the Laplacian $\Delta = - \half (R_+ R_- + R_- R_+ ) = -(R_+ R_- - R_3)$.
The eigenfunctions of $R_+ R_-$ are given by
\begin{align}
u_s &= \sqrt{n+1}\,\bra{\halfn, -\halfn + s} {\hat g} \ket{\halfn, -\halfn} 
= \left[ { (n+1)! \over s! \,(n-s)!}\right]^{\half}
\!\! {z^s \over (1+ \bz z)^{\halfn}}\nonumber\\
\Phi^{(j)}_{m_l} &=  \sqrt{2 j +1}\bra{j, m_l} {\hat g} \ket{j, -\halfn}, \hskip .2in j = {n\over 2} + q, \hskip .1in q = 1, 2, {\rm etc.}
\label{5}
\end{align}
where $\bra{j, m_l} {\hat g} \ket{j, m_r}$ is the representative of the group element $g$ in the spin-$j$ representation of $SU(2)$.
These wave functions are eigenstates of $R_3$ with eigenvalue
$-\halfn$. (Strictly speaking these are sections of a line bundle,
with curvature $\Omega = n d a$; we will refer to them as functions, following the common slight abuse of notation.)
Since $[D_+, D_- ]= - [R_+, R_-] = - 2 R_3$, the combination of curvature and background $U(1)$ field strength is specified by the eigenvalue of $R_3$. For the wave functions in (\ref{5}) with $R_3 = - \halfn$, the background field is that of a magnetic monopole of magnetic charge $n$. There are $N = n+1$ independent functions $u_s$. They are the zero modes of $R_+R_-$, obeying the holomorphicity condition $R_- u_s = 0$. They are also a set of coherent states for $SU(2)$.  One can also view them as the
wave functions obtained in the geometric quantization
of $S^2$ with the symplectic structure $\Omega = n \, da$.
The functions $\Phi^{(j)}_{m_l}$, $j = \halfn + q$, $q = 1, 2, \cdots$, correspond to
the nonzero eigenvalues $n q + q (q+1)$ for $R_+R_-$.
(These eigenfunctions for $R_+ R_-$ for $S^2$ and the generalization to
higher dimensional $\mathbb{CP}^k$ discussed in section 5
are worked out, in the context of the quantum Hall effect, in
\cite{KN-QHE}.)

We now consider the quantum field theory defined by the action
\beq
S = \int dt d\mu \, \left[ i \Psi^\dagger {\del \Psi \over \del t}
- \Psi^\dagger R_+ R_- \Psi \right]
\label{action}
\eeq
The field operator, which will be taken to be fermionic, can be
expanded in terms of the modes in (\ref{5}) as
\beq
\Psi = \sum_{r=1}^N a_r \, u_r + \sum_{j, \alpha} a_{j, {m_l}} \, \Phi^{(j)}_{m_l}
\label{2a}
\eeq
where $a_r$, $a^\dagger_r$ and  $a_{j, {m_l}} $, $a^\dagger_{j, {m_l}} $
form independent sets of canonical fermion
operators, obeying the standard commutation rules
\begin{alignat}{2}
\{ a_r, a^\dagger_s\} &= \delta_{r s},
\hskip .3in
&&\{ a_r, a_s\} = \{ a^\dagger_r, a^\dagger_s\} = 0\nonumber\\
\{ a_{j,{m_l}} , a^\dagger_{j', {m_l}'}\} &= \delta_{j j'} \delta_{{m_l} {m_l}'}
\hskip .3in
&&\{ a_{j,{m_l}}, a_{j',{m_l}'}\} = \{ a^\dagger_{j,{m_l}}, a^\dagger_{j',{m_l}'}\} = 0
\label{18}
\end{alignat} 

The zero modes $u_r$ in (\ref{2a}) will form a basis for the Hilbert space $\H_N$ which is used to define the fuzzy space.
The full Hilbert space of the field theory will contain other states
generated by $a^\dagger_{j, {m_l}}$ as well.
The Berezin-Toeplitz quantization prescription in (\ref{1}) can be rephrased as
\beq
{\hat F}= \int d\mu\, \Psi_0^* \, F(z)\, \Psi_0
= a^\dagger_r \left[ \int d \mu \, u^*_r F(z ) u_s\right] a_s
= a^\dagger_r F_{rs} a_s
\label{2d}
\eeq
where $\Psi_0 = \sum a_r u_r$ denotes the zero-mode part of $\Psi$.
The product of two operators on $\H_N$ can be written as
\beqar
F_{rs} G_{sm} &=& \int d\mu_x u^*_r(x) F(x) u_s (x)  \int d\mu_y
u^*_s(y) G(y) u_m (y)\nonumber\\
&=& \int u_r^* F (z) \left[\sum_s u_s (x) u^*_s(y) \right] G (y) u_m(y)
\label{2f}
\eeqar
We can simplify this further by using the completeness relation
\beq
\sum_s u_s (x) u^*_s(y) = \delta (x, y) -
\sum_{j, \alpha} \Phi^{(j)}_{m_l} (x) \Phi^{(j)*}_{m_l} (y)
\label{2g}
\eeq
If $\M$ is a complex K\"ahler manifold, then the second term on the right hand side can be reduced to
terms with derivatives acting on $\delta (x, y)$ \cite{nair1},
\beqar
\sum_s u_s (x) u^*_s(y) &=& \delta (x, y) +
\sum_k C_k R_+^k R_-^k \, \delta(x, y)\nonumber\\
C_1= - {1\over n+2} , &\hskip .2in& C_2 = {1 \over 2 (n+2) (n+3)}, \hskip .1in
{\rm etc.}
\label{2ga}
\eeqar
(The recursive calculation of the coefficients $C_k$ is given in \cite{nair1}.)
As a result of this relation, (\ref{2f}) can be simplified as
\beq
F_{rs} G_{s m} = \int d \mu\, u^*_r \left( F*G \right) u_m 
\label{2h}
\eeq
where the star product $F*G$ is of the form
\beqar
F*G&=& F G + {1\over n+2}  (R_+F)  (R_-G)  + {1\over 2 (n+2) (n+3)}
(R_+^2 F) R_-^2G) + \cdots\nonumber\\
&=& F\, G + {\rm terms~with~derivatives~of~} F ~{\rm and ~}G
\label{2i}
\eeqar
As is well known, the $*$-product is an associative but noncommutative product for the functions $F$ and $G$. The $*$-commutator has the expansion
\beq
F*G - G*F = {1\over n+2} \Bigl[  (R_+F)  (R_-G) - (R_+G) (R_-F) \Bigr]
+ \cdots
\label{8}
\eeq
The first term on the right hand side is the Poisson bracket
for the symplectic structure $\Omega = n\, d a$.
Combining (\ref{2h}) with
(\ref{2d}), we have
\beq
{\hat F} {\hat G} = \int d\mu\, \Psi^*_0 \, (F*G) \, \Psi_0
= a_r^\dagger \left[ \int d\mu\, u^*_r \, (F*G) \, u_s\right]
a_s
\label{2j}
\eeq
The product of any set of operators on $\H_N$ can be reduced to the form 
in (\ref{2j}) or (\ref{2d}) as a quadratic form
with one power each of $a^\dagger_r$ and $a_s$.

To complete the characterization of the fuzzy space,
we need to specify a state which can be used to associate
a real number to an operator since this is how observables are
related to operators.
We specify $\M_{\rm F}$, the fuzzy version of the manifold
$\M$, as the state
\beq
\ket{\M_{\rm F}} = a^\dagger_{1} 
a^\dagger_{2} \cdots a^\dagger_{N} \, \ket{0}
\label{2k}
\eeq
where $\ket{0}$ denotes the usual vacuum state obeying
\beq
a_i \ket{0} = a_{j, {m_l}} \ket{0} = 0
\label{2c}
\eeq
In this case, the result of the measurement of an operator can be expressed as
\beqar
\bra{\M_{\rm F}} {\hat F} \ket{\M_{\rm F}} &=&
 \bra{0} a_{N} 
\cdots a_2 a_1  \, (a^\dagger_r F_{rs} a_s) \,
a^\dagger_{1} 
a^\dagger_{2} \cdots a^\dagger_{N} \ket{0}
\nonumber\\
&=& \sum_r F_{r r} = \Tr ({\hat F})\nonumber\\
&=& \sum_r\int d\mu\, F(z) \, u_r^* u_r
 \label{2l}
\eeqar

There are a couple of important observations to be made about this state.
If we have a quantum state representing all of space, then the only
value a measurement can return must involve the integral over all space
of the operator. Equation (\ref{2l}) is consistent with this expectation.
If we consider a deformation of the operator $R_+ R_-$ as
$ R_+ R_- \rightarrow R_+ R_- + V$, where $V$ is an operator 
on $\H_N$ of the form
$a^\dagger_r V_{rs} a_s$, we find that the time-evolution of the
state (\ref{2k}) is given by
\beq
e^{- i V t} \ket{\M_{\rm F}} = \det \left( e^{- i V t}\right) \ket{\M_{\rm F}}
= e^{-i (\Tr V)t} \ket{\M_{\rm F}}
\label{time1}
\eeq
This follows from the fermionic nature of the operators $a^\dagger_r$.
Thus, apart from an overall phase, the state $\ket{\M_{\rm F}}$ is invariant, consistent with the fact that we do not expect any nontrivial time-evolution for a closed space
in gravity, beyond just imposing the Hamiltonian
constraint. 
This is to be contrasted with other possible choices of a state.
For example, the state $\rho = (1/N) \sum_r a^\dagger_r \ket{0} \bra{0} a_r$
will also lead to ($1/N$ times ) the trace for an operator
${\hat F} = a^\dagger_r F_{rs} a_s$. However, this state is not preserved by time-evolution, since the commutator
$[ V, a_r^\dagger\ket{0}\bra{0} a_r]$, in general, is not zero.

Given that the state (\ref{2k}) only gives the trace or integral of an operator, we can ask if there is any ``local" observable we can define.
In the absence of matter fields, the only possibility is to consider 
using a set of modified wave functions $\{u'_r \}$, corresponding to some deformation of $R_+ R_-$ or $\Delta$, with
$u'^*_r u'_r - u^*_r u_r$ having support in a compact region of
$\M$. Thus, the only local degrees of freedom associated to gravity must arise from the choices of the wave functions used to construct
the symbols.

If we consider matter fields, with a set of dynamical variables $\{ q_\lambda, p_\lambda \}$, there is an associated operator on $\H_N$
with matrix elements given by \cite{nair3}
\beq
{q}_{rs} = \sum_\lambda q_\lambda \, T^\lambda_{rs}
\label{matter1}
\eeq
where $\{ T^\lambda\}$ form a basis for $N \times N$ matrices.
For simplicity, we take them as hermitian matrices obeying
$\Tr (T^\lambda T^{\lambda'}) = \delta^{\lambda \lambda'}$.
We can also define a contravariant symbol $q (z)$ 
corresponding to ${q}_{rs}$
along the lines of (\ref{1}). If $h(z)$ is a function with support in a compact region $\O_1$, we can construct
\beq
q(h)_{rs} = \int d\mu \, u^*_r (q(z)* h(z)) u_s
\label{matter2}
\eeq
The evaluation of $a^\dagger_r q(h)_{rs} a_s$ on the state
(\ref{2k}) returns the value
\beq
\Tr q(h) = \int d\mu\, q(z)* h(z) \sum_r u^*_r u_r
\label{matter3}
\eeq
The support of $h(z)$ restricts the integration to $\O_1$, so we can consider
(\ref{matter2}) as defining a local operator for matter (nongravitational) fields.

From the mode expansion (\ref{2a}), we have
\beq
a_r = \int d\mu\, u^*_r \Psi , \hskip .3in
a^\dagger_r = \int d\mu\, \Psi^\dagger u_r
\label{II-1}
\eeq
The state (\ref{2k}) can then be written as
\beq
\ket{\M_{\rm F}} = \int d\mu_1 d\mu_2 \cdots d\mu_N\,
u_1(z_1) u_2(z_2) \cdots u_N(z_N) \,
\Psi^\dagger (z_1) \Psi^\dagger (z_2) \cdots \Psi^\dagger (z_N) \ket{0}
\label{II-2}
\eeq
Our second point is that, since this is expressed in terms of local fields
$\Psi^\dagger (z)$, it is possible to use this to define the notion of
entanglement for a fuzzy space. We can ``integrate over" fields in
$\M / \O_1$ to define a reduced density matrix and a corresponding entanglement entropy.
This is explained in the next section.

\section{Defining entanglement}

The procedure for ``integrating out" the fields in some region, and thereby defining a reduced density matrix associated to (\ref{2k}) or (\ref{II-2}),
is most transparent if we use fermionic coherent states. For
operators $a$, $a^\dagger$ obeying the standard anticommutation rules, we can define the coherent states
\beqar
\ket{\theta} &=& \exp\left({\half {\bar\theta}\theta}\right) \, e^{\theta a^\dagger} \ket{0},
\hskip .2in a \ket{\theta} =\theta\, \ket{\theta}\nonumber\\
\bra{\theta} &=& \exp\left({\half {\bar\theta}\theta}\right) \, \bra{0}
e^{{\bar\theta} a}
\label{II-3}
\eeqar
where $\theta$, ${\bar \theta}$ are Grassmann variables.
For a state $\ket{\alpha}$, we also have
\beq
\braket{\theta|\alpha} = \exp\left({\half {\bar\theta}\theta}\right)\,
f_\alpha(\theta)
\label{II-4}
\eeq
The normalization of the states is defined by
\beq
\braket{\alpha|\alpha} = \int d\theta d{\bar\theta}\, \exp\left({{\bar\theta}\theta}\right)~{\bar f}_\alpha f_\alpha
\label{II-5}
\eeq
For the present problem, we have Grassmann variables $\psi$, $\bpsi$
associated to $\Psi$, $\Psi^\dagger$. Equivalently, we can use
Grassmann variables $\theta$, ${\bar \theta}$ for $a_r, a_{j,\alpha}$ and
$a^\dagger_r, a^\dagger_{j, \alpha}$. The canonical structure for these is given by the action (\ref{action}) as
$\omega = i \int d\mu\, \delta \bpsi \delta \psi$.
The coherent states can be defined by
\beqar
\ket{\psi} &=& \exp\left( {1\over 2} \int d\mu\, \bpsi \psi \right)
\, \exp\left( \int d\mu\, \bpsi \Psi^\dagger \right) \ket{0}\nonumber\\
&=&\exp\left(  {\bar \theta}\cdot \theta \right) \,
\exp\left( {\bar\theta} \cdot a^\dagger \right) \ket{0}
\label{II-6}
\eeqar
where ${\bar\theta} \cdot \theta = \sum {\bar\theta}_r \theta_r 
+ \sum {\bar \theta}_{j, \alpha} \theta_{j, \alpha}$, etc.
The corresponding version of the state (\ref{2k}) or (\ref{II-2}) is
given by
\beqar
\ket{\M_{\rm F}}&=& \exp\left({1\over 2}{\bar \theta}_r \theta_r +
{1\over 2} {\bar\theta}_{j,\alpha} \theta_{j, \alpha} \right)
{\bar\theta}_1 {\bar\theta}_2 \cdots {\bar\theta}_N\nonumber\\
&=&e^{\half \int \bpsi \psi} \,{1\over N!} \e^{i_1 i_2 \cdots i_N} \left(\int u_{i_1}\bpsi
\right)
\left(\int u_{i_2}\bpsi\right) \cdots \left(\int u_{i_N}\bpsi\right)
\label{II-7}
\eeqar

Consider separating $\M$ into two disjoint regions $\O_1$, $\O_2$ with a
common boundary and $\O_1 \cup \O_2 = \M$.
We define
\beq
M_{rs} = \int_{\O_1} d\mu\, u^*_r u_s, \hskip .3in
N_{rs} = \int_{\O_2} d\mu\, u^*_r u_s = \delta_{rs} - M_{rs}
\label{II-8}
\eeq
$M$ is an $N \times N$ matrix which is hermitian and positive.
 Diagonalizing it, we write
\beq
M_{rs} = U_{rk} \lambda_k U^\dagger_{ks}, \hskip .2in
N_{rs} = U_{rk} (1- \lambda_k) U^\dagger_{ks}
\label{II-9}
\eeq
where $U$ is unitary and
$0\leq \lambda_k \leq 1$. Using these relations, we write
\beq
u_r = U^*_{r k} ( v_k + w_k)
\label{II-10}
\eeq
where $v_k$ has support in $\O_1$ and $w_k$ has support
in $\O_2$ and
\beqar
\int_{\O_1}  d\mu\, v^*_k v_m &=& \delta_{k m} \lambda_k\nonumber\\
\int_{\O_2}  d\mu\, w^*_k w_m &=& \delta_{k m} (1- \lambda_k)
\label{II-11}
\eeqar
We can now use (\ref{II-10}) to write (\ref{II-7}) as
\beq
\ket{\M_{\rm F}} = e^{\half \int_{\O_1} \bpsi \psi +\half \int_{\O_2} \bpsi \psi} 
(\det U^* )\,\left( \int_{\O_1} v_1 \bpsi + \int_{\O_2} w_1\bpsi \right)
\cdots \left( \int_{\O_1} v_N \bpsi + \int_{\O_2} w_N\bpsi \right)
\label{II-12}
\eeq
(Notice that the state $\ket{|M_{\rm F}}$ in (\ref{2k}) has
antisymmetry under permutation of the labels, this is what allows us to factor
out $\det U^*$ in this equation.)
We now form $\ket{\M_{\rm F}}\bra{\M_{\rm F}}$ and integrate over the
fields in $\O_2$ to obtain the reduced density matrix
$\rho_{\rm Red}$. The key result for the simplification is
\beqar
\int [d\psi d\bpsi]_{\O_2} e^{\int_{\O_2} \bpsi \psi} 
\int_{z, z'} w_i (z) \bpsi (z) w_j^* (z') \psi (z') &=&
\int w_i (z)w_j^* (z') \delta (z, z' ) \nonumber\\
&=& \delta_{ij} (1-\lambda_i)
\label{II-13}
\eeqar
The reduced density matrix takes the form
\beqar
\rho_{\rm Red} &=&e^{\int_{\O_1} \bpsi \psi} \Bigl[
\int v_1 \bpsi \int v_2 \bpsi \cdots \int v_N\bpsi\ket{0} \bra{0} \int v^*_N \psi \cdots
\int v^*_2 \psi \int v^*_1 \psi\nonumber\\
&&+ (1-\lambda_1) \int v_2 \bpsi \cdots \int v_N\bpsi\ket{0} \bra{0} \int v^*_N \psi \cdots
\int v^*_2 \psi \nonumber\\
&&+ (1-\lambda_2)  \int v_1 \bpsi \int v_3 \bpsi\cdots \int v_N\bpsi\ket{0} \bra{0} \int v^*_N \psi \cdots
\int v_3^* \psi \int v^*_1 \psi
+\cdots\Bigr]
\label{II-14}
\eeqar
There are $2^N$ states which occur in $\rho_{\rm Red}$.
Since
\beq
\int [d\psi d\bpsi]_{\O_1} e^{\int_{\O_1} \bpsi \psi} 
\int  v_i  \bpsi \int v_j^* \psi  = \delta_{ij}  \lambda_i 
\label{II-15}
\eeq
the normalization of states in (\ref{II-14}) will produce additional
factors of $\lambda$'s. It is then easy to see that, 
writing $\rho_{\rm Red}$ in terms of normalized states,
it takes the matrix form
\beq
\rho_{\rm Red} = \prod_{\otimes_i} \left[
\begin{matrix}
\lambda_i &0\\
0& (1-\lambda_i) \\
\end{matrix} \right]
\label{II-16}
\eeq
The corresponding von Neumann entropy is thus given by
\beq
S_{\rm EE} = - \sum_{i =1}^N \left[ \lambda_i \log \lambda_i + (1-\lambda_i ) \log (1- \lambda_i )
\right]
\label{II-17}
\eeq
This identifies the entanglement entropy (EE) for the fuzzy space.
A formula similar to this has been obtained for fermionic lattice systems 
\cite{peschel} and has also been used  for calculating the entanglement entropy for
quantum Hall systems \cite{rod-sier}-\cite{nair2}.  It has also been argued to be 
applicable for fuzzy spaces \cite{nair2}.
What is new here is the justification for its use for fuzzy spaces and
its derivation by directly integrating out the fields in one region, say $\O_2$,
by use of fermionic coherent states.

A few remarks are in order at this point. Only the zero mode part
of the field operator $\Psi^\dagger$, either
$a^\dagger_r$ (as in (\ref{2k})) or the integral $\int u_r \Psi^\dagger$
(as in (\ref{II-2})), was used to define
$\ket{\M_{\rm F}}$. However, in the reduced density matrix 
the higher modes $\Phi^{(j)}_{m_l}$ also contribute.
Using the expansion (\ref{2a}),
\beq
\int_{\O_1} u_r \Psi^\dagger =
\sum_r \int_{\O_1} u_r u^*_s \,a^\dagger_s +  \sum_{j, {m_l}} \int_{\O_1}
u_r \Phi^{(j)*}_{m_l}
a^\dagger_{j, {m_l}}
\label{II_18}
\eeq
The integral in the second term, namely, $ \int_{\O_1}
u_r \Phi^{(j)*}_{m_l}$ is not zero, since the range of integration is
only $\O_1$ and so orthogonality is not obtained.
Thus the higher modes of $R_+ R_-$ are important in 
defining a reduced density matrix relevant to local observables.
This is to be expected since localization 
(or defining operators with support in some subregion of $S^2$)
requires all 
eigenfunctions for reasons of completeness. (For example, notice that
the sum $\sum\psi_r (z) \psi^*_r(z')$ is nonlocal on $S^2$, but
$\sum \psi_r (z) \psi^*_r(z') + \sum \Phi^{(j)}_{m_l} (z) \Phi^{(j)*}_{m_l}(z')
= \delta (z, z')$ is local on $S^2$.)
This is also in accord with
the fact that the construction of the $*$-products as in (\ref{2i}) 
requires all eigenfunctions. However, it is worth emphasizing that
no inputs other than the choice of $R_+ R_-$ are needed for this.

\section{Perturbation theory for changes in entropy}

The explicit calculation of entanglement entropy (\ref{II-17})
is mathematically identical to the calculation of the same quantity for
the $\nu =1$ state in quantum Hall effect.
As mentioned earlier, this has been analyzed by many authors and, as expected, the leading term is proportional to the phase volume
(defined by the symplectic structure $\Omega$)
of the interface between $\O_1$ and $\O_2$ \cite{DK}.
These calculations were all done for the case of a fixed background, specified by the K\"ahler two-form $\Omega = n d a$ on $\M$.
For example, for the case of $S^2$, the relevant background is just the
$U(1)$ gauge field $n a$, with $a$ as in (\ref{4}). Our aim is to consider how the EE changes when the background is perturbed, as in
$a \rightarrow a +A$ for $S^2_{\rm F}$. Since the EE is determined by
$\lambda_s$, this is equivalent to asking how 
$\lambda_s$ change under $a \rightarrow a +A$. We will work this out
explicitly for the case of $S^2_{\rm F}$ using perturbation theory.
A more general argument applicable to higher dimensional cases
will be given in the next section.

Using the perturbed background $a +A$
is equivalent to using 
$(R_+ + A_+) (R_- + A_-)$ in place of the operator $R_+ R_-$.
Towards using perturbation theory for this, consider $U^\dagger (R_i +A_i) U$, with
\beq
U = e^{i H}, \hskip .2in H =  R\cdot \xi + \xi\cdot R + \cdots
\label{9}
\eeq
where $\xi_i$ is to be determined.
(Caution: $U$ here is a unitary operator different from the $U$
with matrix elements $U_{rs}$ given in (\ref{II-9}).)
We then find
\beq
U^\dagger R_i U = R_i - 2 \epsilon_{ij3} \xi_j R_3 + i \left[ R_j [ R_i, \xi_j] 
+ [R_i, \xi_j] R_j + i \epsilon_{ij3} [R_3, \xi_j] \right] + \cdots
\label{10}
\eeq
The commutator $[R_i, \xi_j]$ involves gradients of $\xi_j$. We will consider an expansion in gradients of $A_i$ as well as powers of $A_i$, so that this term can be considered as being of higher order compared to
$\epsilon_{ij3} \xi_j R_3$. By virtue of the Jacobi identity, we have the relation
\beq
\epsilon_{ij3} [R_3, \xi_j] = \epsilon_{ij3}\, \epsilon^{pq 3}\, [R_p, [R_q,\xi_j]]
\label{11}
\eeq
This shows that $\epsilon_{ij3} [R_3, \xi_j] $ has two derivatives of $\xi_j$, accordingly we have grouped this term along with the derivative terms in
(\ref{10}). We then have
\beq
U^\dagger (R_i + A_i) U \, u_s =
R_i \, u_s  + ( n \epsilon_{ij3} \xi_j + A_i ) \, u_s + \cdots
\label{12}
\eeq
where the ellipsis indicates terms of order $\xi^2$ or terms involving derivatives of $\xi$.
We can now choose $\xi_i$ to the first order in $A_i$ as
$\xi_i = \epsilon_{ij3} A_j/n$, so that, to the order calculated,
\beq
U^\dagger (R_i +A_i ) \,U\, u_s \approx R_i \, u_s
\label{13}
\eeq
This shows that, again to the order calculated, the lowest level eigenfunctions of $(R_+ +A_+) (R_- + A_-)$ can be taken to be
\beqar
\phi_s &\approx& U \, u_s \approx \left[1 + {i \over n} \epsilon_{ij3} (R_i A_j + A_j R_i) \right] u_s \nonumber\\
&\approx& \left[1 + { 2 i \over n} \epsilon_{ij3}  A_j R_i \right] u_s +\cdots
\label{14}
\eeqar
These functions also obey the modified holomorphicity condition
$(R_- + A_- ) \phi_s \approx 0$. The symbols are now given in
terms of these eigenfuncctions, as in (\ref{1}),
\beq
F_{rs} = \int d\mu \, \phi^*_r \, F (\bz, z) \, \phi_s 
= \int d\mu\, u^*_r (U^\dagger F U)\, u_s
\label{15}
\eeq
The $*$-product now takes the form
\beq
F*G = F\, G +{1\over n+2} (R_+ +A_+)F\, (R_- +A_-) G + \cdots
\label{16}
\eeq
This shows clearly that the classical limit (e.g. the Poisson bracket)
obtained as $N \rightarrow \infty$ is different for different choices of $A_\pm$.

For simplicity, consider the case of $\O_1$ having azimuthal symmetry,
so that $\int_{\O_1} u^*_r u_s$ is nonzero only for $r =s$ by the angular
part of the integration. For the unperturbed  case, we then have
$\lambda_s\vert_{A =0} = \int_{\O_1} u^*_s u_s$.
Using $\phi_s$ from (\ref{14}), we find
\beqar
\lambda_s&=& \int_{{\cal O}_1} \left( u^*_s + {2\over n } A_j \epsilon_{ij3} D_i u^*_s\right) \left( u_s + {2\over n } A_b \epsilon_{ab3} D_a u_s\right) \nonumber\\
&\approx& \lambda_s \vert_{A=0} + {2\over n} \int_{{\cal O}_1}\epsilon_{ij3} A_j \del_i (u^*_s u_s)
+\cdots\nonumber\\
&\approx&  \lambda_s \vert_{A=0} + {2\over n} \int_{{\cal O}_1} d(u^*_s u_s)\,A  +\cdots
=  \lambda_s \vert_{A=0} + {2\over n} \oint_{\del {\cal O}_1}  u^*_s u_s \, A +\cdots
\label{28}
\eeqar
In the last line we used the notation of differential forms. In addition to the boundary term displayed, one does get a term proportional to $dA$. But, since we have not systematically evaluated terms involving derivatives of
$A$, this term must be considered as being of higher order and
has been absorbed as part of the terms indicated by the ellipsis.
The result (\ref{28}) shows that $\delta \lambda_s$ is proportional to the 
one-dimensional Chern-Simons form $A$ integrated (with the additional factor
$u^*_s u_s$) along the boundary of ${\cal O}_1$. Thus
\beq
\delta S = - {2\over n} \oint_{\del {\cal O}_1}  \!\! A \left[
\sum_s (u^*_s u_s)\vert_{\del{\cal O}_1} \log \left( {\lambda_s \over 1- \lambda_s}\right)\right] +\cdots
\label{29}
\eeq
This is a key result. (This result was already given in \cite{nair2},
with a different set of arguments leading to it. 
The novel result here is a direct calculation using perturbation theory.)
Since we did this calculation for two dimensions and using perturbation theory, the generality of the result may not be clear.
We will now turn to the more general situation.

\section{Entanglement for general fuzzy spaces}

In this section, we will consider the generalization of the result for
$S^2_F$ to fuzzy spaces in higher dimensions. 
We consider the case of fuzzy $\mathbb{CP}^k$; this will provide the
simplest generalization.
For our purpose, 
${\mathbb{CP}^k}$, $k > 1$
may be viewed as a group coset,
\beq
{\mathbb{CP}^k} = {SU(k+1) \over U(k)}
\label{30}
\eeq
The Riemannian curvatures take values in the Lie algebra of
$U(k)$ and are constant in the tangent frame basis.
Again, we consider a symplectic form which is an integer multiple of the
K\"ahler form. Geometric quantization then leads to the Hilbert space
$\H_N$ which defines the fuzzy version of $\mathbb{CP}^k$.
In the holomorphic polarization, we get a set of holomorphic 
sections of the line bundle which we may refer to as the
wave functions.
Since $U(k)$ is the structure group of $SU(k+1)$ as a bundle over
$\mathbb{CP}^k$, it is possible to consider more general
$\underline{U(k)}$-valued connections. 

As in the previous sections, we want to consider
$\H_N$ in terms of the eigenfunctions of the
Laplace operator. Towards this, let $g$ be a matrix representative of an
arbitrary element in the group $SU(k+1)$, say, in the fundamental
representation. We choose an orthonormal basis of matrices $\{ T_A\}$ for the Lie algebra of $SU(k+1)$. This can be
further split as
\beq
\{ T_A \} = \{ T_a, T_{k^2 + 2k}, T_{+i} , T_{-i} \}, \hskip .2in
a = 1, 2,\cdots, (k^2-1); \hskip .1in i = 1, 2, \cdots, k
\label{31}
\eeq
where $\{ T_a \} \in \underline{SU(k)}$, $T_{k^2 + 2k} \in \underline{U(1)}$
form a basis for the Lie algebra of the $U(k)$ subgroup
and $T_{\pm i}$ correspond to the coset directions.
We define right translation operators on $g$ by
\beq
R_{\pm i} \, g = g\, T_{\pm i}, \hskip .2in
R_a \,g = g\, T_a, \hskip .1in
R_{k^2+ 2k} \,g = g\, T_{k^2 + 2k}
\label{32}
\eeq
The Laplace operator $\Delta$ on functions is given by
\beq
- \Delta = {1\over 2} \left( R_{+i} R_{-i} + R_{-i} R_{+i} \right)
= R_{+i} R_{-i} - \sqrt{2k (k+1)}\, R_{k^2 + 2k}
\label{33}
\eeq
The eigenfunctions of $\Delta$ are given in terms of
unitary irreducible representations of $SU(k+1)$; these are also finite dimensional.
We denote the representative matrices for $g$ by
\beq
\D^{(J)}_{{m_l}; {m_r}}(g) = \la J ,{m_l}\vert\, g\,\vert J, {m_r} \ra \label {34}
\eeq
where $J$ labels the specific irreducible representation
and ${m_l}, ~{m_r} $ stand for two sets of quantum numbers specifying the states within the representation.
The eigenfunctions of $-\Delta$, which we will denote by
$u_{m_l}$, obey the condition
\beqar
R_{k^2 + 2k} \, u_{m_l} &=&  - {n k \over \sqrt{2 k (k+1)}} \, 
u_{m_l}\nonumber\\
R_a \, u_{m_l} &=& 0
\label{35}
\eeqar
These conditions ensure that $u_{m_l}$ are sections of a scalar bundle
(no spin) with $U(1)$ curvature equal to $n$ times the K\"ahler form 
$da$ and $SU(k)$ curvature
equal to zero. With proper normalization, these wave functions can be taken to be
\beq
u_{m_l} = \sqrt{{\rm dim}J} \, \D^{(J)}_{{m_l}; w}
= \sqrt{{\rm dim}J}\, \la J ,{m_l}\vert\, g\,\vert J, w \ra
\label{36}
\eeq
where $\ket{J, w}$ is chosen to ensure the conditions
(\ref{35}). The lowest eigenfunctions of $-\Delta$ obey the
additional condition
\beq
R_{-i}\, u_s = 0
\label{37}
\eeq
This is in agreement with the form of $-\Delta$ in (\ref{33}) and
is the holomorphicity condition consistent with the geometric
quantization of $\Omega = n\, d a$.

The lowest eigenfunctions (\ref{37}) correspond to
the rank $n$ totally symmetric representations
of $SU(k+1)$. Explicitly using complex coordinates
for ${\mathbb{CP}^k}$, we can write them out as
\beq
u_s = \sqrt{N}\,
\left[ {n! \over i_1! i_2! \cdots i_k! (n-l)!}\right]^{\half}\,
{z_1^{i_1} z_2^{i_2}\cdots z_k^{i_k} \over (1 + \bz\cdot z)^{n \over 2}}
\label{38}
\eeq
Here
$l = i_1 + i_2 +\cdots+ i_k$ and $N = {\rm dim}J = (n+k)!/ (n! k! )$ is the 
total number of states. This will be the dimension
of $\H_N$ for the fuzzy version of
$\mathbb{CP}^k$.
The index
 $s$ takes values $1, 2, \cdots, N$, corresponding to the possible choices
of the set of numbers
$(i_1, i_2, \cdots i_k)$, starting with $s= 1$ for
$(0,0, \cdots, 0)$. 
For the normalization of $u_s$ in (\ref{38}), we have used
the volume element for $\mathbb{CP}^k$, normalized to $1$; i.e., 
\beq
d\mu = {k! \over \pi^k} {d^2z_1 \cdots d^2z_k
\over (1+ \bz \cdot z)^{k+1}}, \hskip .2in \int d\mu = 1
\label{39}
\eeq

Now that we have obtained $u_s$, the calculation of the entanglement entropy can proceed along similar lines to what was done in the last section. 
If $\O_1$ has azimuthal symmetry, we can write
\beq
\lambda_s  = \int_{\O_1} d\mu \,u^*_s u_s
\label{40}
\eeq
Using (\ref{38}) for $u_s$ in this equation and then using
the $\lambda_s$ in  (\ref{II-17}), we can calculate the entanglement entropy
for background values of fields which correspond to
the spin connection $A^{\rm sp}_0$ for the Fubini-Study metric,
with an additional $U(1)$ connection $n a$ (which defines $\Omega$).
Our aim is to include perturbations of this background.
It is possible to use perturbation theory as we did in section 3 and
obtain a result very similar to (\ref{28}) or (\ref{29}).
However, we will now present a different argument
which helps to treat the more general case. 
(In this section we use results from \cite{KN-index} and \cite{nair2}
with some clarifications and elaborations.)

We start with two important
ingredients which are needed before this argument can be formulated.
The first result we need is the Dolbeault index density.
Because of the holomorphicity  condition (\ref{37}) or 
its generalization to arbitrary background
connections, the wave functions which define $\H_N$ belong
to the kernel of 
$\bdel$-operator. Therefore the number of states 
is given by the Dolbeault index theorem \cite{KN-index}.
This theorem states that the index is given as
\beq
{\rm Index} (\bdel_V) = \int {\rm td}(T_c K) \wedge {\rm ch}(V)
\equiv \int {\cal I}_{\rm Dolb}
\label{41}
\eeq
where ${\rm td}(T_cK)$ is the Todd class on the complex tangent space of
the manifold $K$
and ${\rm ch}(V)$ is the Chern character of the vector bundle $V$.
The vector bundle refers to the fact that we can have 
gauge fields in addition to the gravitational fields.
$K$ is topologically $\mathbb{CP}^k$ but the connection on the vector
bundle as well as the spin connection are more general.
The Todd class is given by
\begin{align}
{\rm td} &= \prod_i {x_i \over 1- e^{-x_i}}\nonumber\\
&= 1 + {1\over 2} \, c_1 +{1\over 12} ( c_1^2 + c_2) + {1\over 24} c_1\, c_2
+ {1\over 720} ( - c_4 + c_1\,c_3 + 3 \, c_2^2 + 4\, c_1^2 \,c_2 - c_1^4) + \cdots
\label{42}
\end{align}
The splitting principle gives the formula as written in the first line;
in the second line, we give the expansion for low dimensions in terms of the Chern classes.
Just for completeness, the Chern classes $c_i$ and the Chern character for the vector bundle with curvature ${\cal F}$
are given by
\beqar
\sum_i c_i \, t^i &=&\det \left( 1 + {i \, {\cal F} \over 2 \pi} \,t\right) \nonumber\\
{\rm ch}(V) &=& \Tr \left( e^{i {\cal F} /2 \pi} \right) = {\rm dim}\,V + \Tr ~{{i {\cal F}} \over {2\pi}} + { 1 \over 2!} \Tr~ {{i{\cal F} \wedge i {\cal F}} \over {(2\pi)^2}} + \cdots
\label{43}
\eeqar
The Dolbeault index, for low dimensions, has
the expansion
\beqar
{\rm Index} (\bdel_V)
\!\! &=&\!\! \int {\rm dim} V\, \Tr \left({i R \over 4\pi}\right)  + \Tr \left( {i F \over 2 \pi}\right)
+ {{\rm dim}V \over 12} ( c_1^2 + c_2) + {1\over 2} \Tr \left( {i F \over 2 \pi} \right)^2
+ \cdots\nonumber\\
c_1 \!\!&=&\!\!\Tr {i R \over 2 \pi} , \hskip .2in
c_2 = {1\over 2} \left[ \left( \Tr {i R \over 2 \pi}\right)^2 - \Tr \left({i R \over 2 \pi}\right)^2
\right]
\label{44}
\eeqar
where, as usual, one has to pick the terms which are $2k$-forms 
for results pertinent to
$2k$ real dimensions.

The second result we need is a generalized Chern-Simons form.
The index density ${\cal I}_{\rm Dolb} \equiv {\rm td}(T_c K) \wedge {\rm ch}(V) $ from
(\ref{41})  defines an invariant polynomial of the field
strength $F$ and the curvature $R$.
We can 
now define a generalized Chern-Simons term
corresponding to ${\cal I}_{\rm Dolb} (\F) $ by the formula
\beq
Q(A_2, A_1)= k \int_0^1 dt~ {\cal I}_{\rm Dolb} (A_2 -A_1, \F_t, \F_t, \cdots, \F_t),
\hskip .2in A_t = A_1 + t (A_2 -A_1)
\label{45}
\eeq
where $\F$ can be $F$ or $R$, with $A$ referring to the gauge connection or the spin connection, respectively. 
Terms in the index density are traces of symmetric polynomials of
$\F$, so ${\cal I}_{\rm Dolb} (A_2 -A_1, \F_t, \F_t, \cdots, \F_t)$
stands for replacing one of the $\F$'s in each monomial by
$A_2 - A_1$, while the rest are taken to be $\F_t$.
It is easy to verify that $Q(A_2, A_1)$ obeys
\beq
 {\cal I}_{\rm Dolb} (\F_2) - {\cal I}_{\rm Dolb} (\F_1) = d \, Q(A_2, A_1)
\label{46}
\eeq
The connections $A_2$ and $A_1$ belong to the same topological class in the sense that the integrated index is the same for both, and one can continuously connect $A_2$ to $A_1$
as in $A_t = A_1 + t (A_2 -A_1)$.
So we may regard $A_2$ as a perturbation of $A_1$.

The starting background has
the spin connection $A^{\rm sp}_0$ given by the Fubini-Study metric
and the $U(1)$ background field $\Omega$ proportional to
the K\"ahler form $d a$.
The perturbations $a \rightarrow A$ and $A^{\rm sp}_0 \rightarrow A^{\rm sp}$ 
will modify the wave functions $u_s$. Let $\phi_s$ denote these modified
wave functions. The use of these will lead to modified
symbols as explained in the introduction, and this, in turn, defines the large $N$ limit (as characterized by the data $(A, A^{\rm sp})$).
Towards identifying the dependence of the EE on this background data, we define a diagonal matrix $P (s)$
by $P_{ii} = 1$ for $i = 0, 1, \cdots, s$, with all other elements equal to zero.
Associated to this, we can define a function
\beq
P(s) = \sum_{k= 1}^{s}  \phi_k \phi^*_k
\label{47}
\eeq
This function is the covariant symbol as defined in (\ref{2}) for the matrix $P$ apart from the normalization factor $\C$. ($\C$ is actually equal to
$N$, but it will not be important for the present argument.)
The reason behind looking at $P$ is that $P (s) - P(s-1) = \phi_s \phi_s^*$.
In the following, let $P$ denote this function when
$\phi$'s correspond to $(A, A^{\rm sp})$, and $P_0$ denote the same for
the starting connections $(a, A^{\rm sp}_0 )$.
We also define $^*P$ as the $2k$-form dual to $P$ (which also uses the volume form 
with the background $(A, A^{\rm sp})$).
On general grounds, we will see that $^*P$ has to be of the form
\beq
^*P = {{\cal I}_{\rm Dolb} \over N} P_0 - K \, dP_0 + d \mathbb{X}
\label{48}
\eeq
Our strategy will be to motivate and determine the form of
various terms in this equation by considering different special cases.

Towards this, consider the case where 
$s= N$, so that all states in $\H_N$ are used in
$^*P$, $P_0$. In this case, $P_0$ is a constant
since
\beq
P_0 = \sum_{1}^N u_k u^*_k = N \, \D_{k, w} \D^*_{k, w} = N \D_{w, w} (g^\dagger g )
= N
\label{49}
\eeq
In this case, we find from
(\ref{48}) that $^*P = {\cal I}_{\rm Dolb} + d \mathbb{X}$, consistent
with the integral of $^*P$ being given by the index theorem.

Another special case to consider is when the perturbations are zero,
but we keep $s  < N$.
In this case, we should have $P = P_0$, although $P_0$ will not be a constant since the summation over $k$ in (\ref{49}) does not
include all states.
Further, since the curvatures are all constant when the perturbations are zero,
${\cal I}_{\rm Dolb}$ should be a constant times the volume form (\ref{39}).
Since $\int {\cal I}_{\rm Dolb} = N$, we see that we must have
${\cal I}_{\rm Dolb}(a, A^{\rm sp}_0)  = N d \mu$. We see that we do obtain 
$P = P_0$ from (\ref{48}) provided $K$ and $d \mathbb{X}$ vanish
for $(a, A^{\rm sp}_0)$.
Further, since $\lambda_s $ should be $  \int_{{\cal O}_1} P_0 d\mu$ for this case, we also see that we need 
\beq
\oint_{\del{\cal O}_1} \mathbb{X} = 0
\label{50}
\eeq
for any choice of ${\cal O}_1$. Thus we need $\mathbb{X}$, not just
$d\mathbb{X}$ to vanish when the perturbations are zero.

Finally, consider again
the case $s < N$, but with perturbations included.
We expand $\I_{\rm Dolb}$ around $(a, A^{\rm sp}_0)$ 
using (\ref{46}) to get
\beq
\I_{\rm Dolb}  (a+A, A^{\rm sp}) = \I_{\rm Dolb} (a, A^{\rm sp}_0 ) + d Q 
= N \,d\mu + d Q
\label{51}
\eeq
Using this relation, (\ref{48}) becomes
\beq
^*P = d\mu \,P_0 + {d Q\over N} \, P_0 - K \, d P_0 + d {\mathbb X}
\label{52}
\eeq
Since the total number of states should be the same for both
$(a, A^{\rm sp}_0 )$ and $(a+A, A^{\rm sp})$, terms
in (\ref{52}) other than $d\mu\, P_0$
must combine into a total derivative, so that they can give zero
upon integration. This identifies
$K = Q/N$. The final formula for $^*P$ then takes
the form
\beq
^*P(s)  = d\mu\, P_0(s)  + d \left( {Q P_0(s) \over N}\right)
+ d \mathbb{X(s)}
\label{53}
\eeq
The $s$-dependence of $P$, $P_0$ is shown explicitly again in this equation.
As mentioned earlier, $\mathbb{X}$ must vanish when
the perturbations are zero. Our arguments so far do not 
completely fix the form
of $\mathbb{X}$.

Integrating (\ref{53}) over ${\cal O}_1$ and taking the difference for 
$s$ and $s-1$, we get
\beq
\lambda_s (a+A, A^{\rm sp}) =
\lambda_s (a, A^{\rm sp}_0) + {1\over N} \oint_{\del {\cal O}_1}  Q(a+A, A^{\rm sp}; a, A^{\rm sp}_0) \,u^*_s u_s + \oint_{\del {\cal O}_1}  \left[ \mathbb{X} (s) - \mathbb{X} (s-1)\right]
\label{54}
\eeq
Using the difference of the $\lambda_s$'s as given by this
expression, we obtain the field dependence
of the entropy, to first order in $Q$, $\mathbb{X}$ as
\beqar
\delta S &=& - {1\over N} \oint_{\del {\cal O}_1} Q(a+A, A^{\rm sp}; a, A^{\rm sp}_0) \,\sum_s (u^*_s u_s)\vert_{\del{\cal O}_1} \log \left({ \lambda_s \over
1- \lambda_s} \right) \nonumber\\
&& - \oint_{\del {\cal O}_1} \sum_s \bigl[ \mathbb{X}(s) - \mathbb{X}(s-1)\bigr]_{\del{\cal O}_1}
\log \left( {\lambda_s \over
1- \lambda_s} \right) 
\label{55}
\eeqar
This is the generalization of the result we obtained for the
two-dimensional case in section 4.
For the two-dimensional case, the Dolbeault index density
is $(2 F + R )/4\pi$, $N = n+1$, so that for large $N$ the first term on the right hand side of (\ref{55}) agrees with (\ref{29}), apart from a factor of
$4\pi$. This is due to our normalization of the total volume of the manifold being one.
(Notice that if $F/2\pi$ (from the index) integrates to $n$, then, with the volume of $S^2$ taken to be one, it is $2 F$ which integrates to $n$. 
Thus we must multiply $Q$ from the index density by $4\pi$
to get the result in ({\ref{29}).)

The main conclusion of this section, based on
(\ref{55}) is that the change in the entropy is proportional 
to the generalized Chern-Simons form associated with the Dolbeault index.
Since $u_s^* u_s$ and factors involving $\lambda_s$ in (\ref{55}) are calculated for the unperturbed background
$(a, A^{\rm sp}_0)$, the dependence on the
fields is basically given by $ Q(a+A, A^{\rm sp}; a, A^{\rm sp}_0)$ and
$\mathbb{X}(s) - \mathbb{X}(s-1)$. Explicit calculation, at least for some examples, show that $\mathbb{X}(s) - \mathbb{X}(s-1)$ is subdominant,
in the sense of involving gradients of $u^*_s u_s$, compared to
the Chern-Simons term. (See Appendix B of \cite{nair2}.)

As stated in the introduction, one of the motivations for
seeking the field dependence of the EE was to see if one could make 
an argument for optimization of the large $N$ limit as
leading to the field equations for gravity, parallel to what happens with the BH entropy and the Einstein field equations for 
4d gravity \cite{jacobson}.\footnote{Since
the state (\ref{2k}) can also be viewed as the $\nu =1$ 
quantum Hall state, the
calculation of the field dependence of the EE can be of independent interest in that context as well.}
In the context relevant to our analysis in this paper, the
result (\ref{55}) in terms of 
the generalized Chern-Simons term is strongly suggestive of
obtaining CS-type gravity for the $2k +1$ dimensional spacetime
we have, as $N \rightarrow \infty$.

\section{Summary}

The main thrust of this paper has been to contextualize and clarify
some of the ideas and results we have obtained previously. 
We give a definition of a fuzzy space as a state in a quantum field theory
for the degrees of freedom relevant to space. 
This state, we argue, suffices to obtain the results of measurements
for any observable relevant to fuzzy geometry.
In turn, this leads to a
definition and a framework for the entanglement entropy for such spaces.
Recall that, for example, for a free massless
scalar field theory on Minkowski space, we can construct the vacuum wave function as a functional of the fields; it is of the form
$\exp\left( - \half \int \vf (x) \sqrt{-\nabla^2}_{x,y} \vf (y)\right)$. 
In the corresponding density matrix, we can integrate
over the fields
in some region to define a reduced density matrix. The fact
that $\sqrt{-\nabla^2}$ is a nonlocal operator is what leads to
entanglement. 
Once we have the fermionic field theory for a fuzzy space
with a
specific choice of the state,
it is possible to follow a similar strategy and carry out direct integration over the fields in a certain region 
to define a reduced density matrix and 
the corresponding entropy. (This gives an independent derivation
of a formula that has been used in the context of the quantum Hall effect
as well.)
We further consider the change in the EE due to perturbations in the
Laplace operator used to define the fuzzy space. We argue that this is related to a generalized Chern-Simons form
corresponding to the Dolbeault index density.
This makes a point of contact with the entropic approach to gravity,
specifically, in the present case,
for Chern-Simons gravities in odd spacetime dimensions.

\bigskip

I thank Dimitra Karabali for a critical reading of the manuscript. 
This work was supported in part by the U.S. National Science Foundation Grants No. PHY-2112729.

%%%%%%%%%%%%%%%%%%%%%%%%%%%%%%%%%%%%%%%%%%%%%%%%

%%%%%%%%%%%%%%%%%%%%%%%%%%%%%%%%%%%%%%%%%%%%%%%%
%%%%%%%%%%%%%%%%%%%%%%%%%%%%%%%%%%%%%%%%%%%%%%%%
%%%%%%%%%%%%%%%%%%%%%%%%%%%%%%%%%%%%%%%%%%%%%%%%
%%%%%%%%%%%%%%%%%%%%%%%%%%%%%%%%%%%%%%%%%%%%%%%%

%%%%%%%%%%%%%%%%%%%%%%%%%%%%%%%%%%%%%%%%%%%%%%%%
%%%%%%%%%%%%%%%%%%%%%%%%%%%%%%%%%%%%%%%%%%%%%%%%
%%%%%%%%%%%%%%%%%%%%%%%%%%%%%%%%%%%%%%%%%%%%%%%%
%%%%%%%%%%%%%%%%%%%%%%%%%%%%%%%%%%%%%%%%%%%%%%%%
\end{document}